Виктор Кромер, Лев Кромер

**Правильно ли мы считаем калории?**

Наука считать калории медленно, но верно овладевает широкими массами. Этому учат в школе на уроках биологии, умение считать калории популяризуется сторонниками здорового образа жизни, и без подсчета калорий никак не обходятся ни врач-диетолог, составляющий рацион для пациента с избыточным весом, ни дежурный по воинской части, проверяющий правильность дневного расклада продуктов.

Суть теории о калориях лучше всего изложена противником этой самой теории, Мишелем Монтиньяком. "Человеческому телу нужна энергия. Прежде всего, чтобы поддерживать температуру на уровне 36,5 °C. Как только человек начинает двигаться, появляется потребность в энергии. А чтобы переваривать пищу, требуется еще больше энергии. Потребности в энергии меняются в зависимости от возраста, пола и профессиональной деятельности" [1, с, 27].

Суть теории изложена совершенно верно, но, к сожалению, опровергнуть ее походя, как это делает Монтиньяк, невозможно. Существуют достаточно простые химические и физические соотношения, и никакому биологическому объекту нарушить эти законы не дано – эти законы существовали еще до появления биологических форм. Отказать в последовательности Монтиньяку нельзя – отвергая влияние избыточных или недостаточных калорий на процесс набора или сгонки веса, он не признает такого-же действия и за физической активностью ("регулярные занятия спортом вызывают в организме потребность в восполнении затрат, на которую организм отвечает, закладывая резервы гликогена – взамен израсходованного во время ваших упражнений"). Так откуда браться этим новым резервам взамен израсходованных, как не из еды? Значит, при занятиях спортом либо будешь худеть при прежнем рационе, либо нужно будет больше есть, чтобы сохранить прежний вес.

Опровергает Монтиньяк не теорию калорий (так он называет теорию сбалансированного питания), а примитивную версию этой теории, излагаемую с целью якобы популяризации здорового образа жизни. Именно в рамках этой популяризации из пособия в пособия кочуют таблицы, где поясняется, что если вы съели морковку, то для расходования полученной энергии необходимо 8 минут ходить, а после жареного пирожка необходимо 8 минут уже бежать. Можно, впрочем, и 116 минут пассивно отдыхать.

Вот суть популяризируемых знаний: при сжигании в организме 1 грамма белков или углеводов выделяется 4 ккал энергии, а при сжигании 1 г жира – 9 ккал. Эти цифры узаконены и согласно им подсчитывается и печатается на этикетках калорийность продаваемых продуктов питания, исходя из химического состава продукта. Данные сводятся в таблицы и пробегая взглядом по колонке, где, например, указано содержание белков (или жиров, или углеводов), можно быстренько навскидку подобрать продукт с высоким (или низким) содержанием тех-же белков, жиров или углеводов. При этом совершенно упускается из виду, что помимо белков, жиров и углеводов, в продуктах содержится еще, как минимум, вода, и сравнивать нужно не количество белков (жиров, углеводов) в общей массе продукта, а количества белков (жиров, углеводов) в твердом усваиваемом остатке.

И в таблицах типа "Содержание белка в 100 г съедобной части продукта" [2, с. 14], кефир отнесен в группу с малым содержанием белка, а рис – с умеренным, хотя подсчет показывает, что в нежирном кефире на долю белков приходится 43% калорийности продукта, а в рисе – всего лишь 8%. В аналогичной таблице по углеводам виноград отнесен в группу с умеренным содержанием углеводов, а горох – с большим, хотя в винограде доля углеводов в общей калорийности составляет 96%, а в горохе – 66%. (Подсчеты пока ведутся в рамках традиционного подхода). Подобного рода примеры можно множить. Приходим к выводу, что о ценности продукта в отношении содержания в нем определенного пищевого вещества (ингридиента) необходимо судить по доле калорийности этого вещества в общей калорийности, тем самым из расчета исключаются содержащиеся в продукте неусвояемые (балластные) вещества и вода.

Другой источник погрешности связан с неучетом т.н. специфически динамического действия пищи (СДДП). На переваривание, всасывание, транспорт и усвоение питательных веществ пищи требуется энергия, которую взять можно опять-таки только у пищи. Но у белков на СДДП расходуется до 30–40% их энергетической ценности, а у жиров и углеводов лишь 5–7% (это как в дровах – энергия дров идет не только на прогрев печки, но и на испарение содержащейся в дровах-же воды). И напрашивается еще одно уточнение общепринятой методики подсчета энергетической ценности продукта – ценность 1 г жиров по-прежнему считаем за 9 единиц, углеводов за 4 единицы, а вот белков – за 3 единицы. (5–7% снижения энергетической ценности жиров и углеводов мы учли, считая СДДП белков в точности равной 25%). И считать энергетическую ценность продуктов в килокалориях с учетом фактора СДДП некорректно. Назовем предлагаемую новую единицу ЦЕП – ценностная единица продукта.

Ниже приводится таблица энергетической ценности некоторых продуктов в ЦЕП. Химический состав продуктов приводится согласно [2, с. 277–187]. Таблица может быть расширена исходя из химического состава продукта: 1 г белка = 3 ЦЕП, 1 г жира = 9 ЦЕП, 1 г углеводов = 4 ЦЕП. Энергетическая ценность ингридиентов, впрочем, разнится и среди белков, жиров и углеводов в зависимости от их химической формулы, но это уже погрешность второго плана по сравнению с устраняемой нами самой важной погрешностью, связанной с неучетом СДДП белков.

Таблица 1. Энергетическая ценность продуктов

| Продукт | Белки, г | Жиры, г | Углеводы, г | Энергетическая ценность, ЦЕП |
| --- | --- | --- | --- | --- |
| Говядина 1 категории | 18,6 | 14,0 | 0,0 | 182 |
| Печень говяжья | 17,9 | 3,7 | 0,0 | 87 |
| Баранина 1 категории | 15,6 | 16,3 | 0,0 | 194 |
| Свинина мясная | 14,3 | 33,3 | 0,0 | 343 |
| Кролик 1 категории | 21,1 | 15,0 | 0,0 | 198 |
| Куры 1 категории | 18,2 | 18,4 | 0,0 | 220 |
| Сосиски говяжьи (молочные) | 12,3 | 25,3 | 0,0 | 265 |
| Сардельки говяжьи | 9,5 | 17,0 | 1,9 | 189 |
| Колбаса любительская вареная | 12,2 | 28,0 | 0,0 | 289 |
| Колбаса докторская | 12,8 | 22,8 | 0,0 | 244 |
| Яйцо куриное | 12,7 | 11,5 | 0,7 | 144 |
| Камбала | 15,7 | 3,0 | 0,0 | 74 |
| Ледяная рыба | 15,5 | 2,7 | 0,0 | 71 |
| Минтай | 15,9 | 0,9 | 0,0 | 56 |
| Окунь морской | 18,2 | 3,3 | 0,0 | 84 |
| Судак | 18,4 | 1,1 | 0,0 | 65 |
| Треска | 16,0 | 0,6 | 0,0 | 53 |
| Икра зернистая | 31,6 | 13,8 | 0,0 | 219 |
| Морская капуста мороженая | 0,9 | 0,2 | 3,0 | 17 |
| Кальмар | 18,0 | 0,3 | 0,0 | 57 |
| Молоко пастеризированное | 2,8 | 3,2 | 4,7 | 56 |
| Молоко обезжиренное | 3,0 | 0,1 | 4,7 | 28 |
| Кефир жирный | 2,8 | 3,2 | 4,1 | 54 |
| Кефир нежирный | 3,0 | 0,1 | 3,8 | 25 |
| Сметана 10% | 3,0 | 10,0 | 2,9 | 111 |
| Сметана 15% | 2,6 | 15,0 | 3,6 | 157 |
| Сметана 20% жирности | 2,8 | 20,0 | 3,2 | 201 |
| Сметана 35% жирности | 2,0 | 35,0 | 2,6 | 331 |
| Творог жирный | 14,0 | 18,0 | 2,85 | 215 |
| Творог полужирный | 16,7 | 9,0 | 1,3 | 136 |
| Творог нежирный | 18,0 | 0,6 | 1,85 | 67 |
| Сыр голландский | 26,8 | 27,3 | 0,0 | 326 |
| Мороженое сливочное | 3,3 | 10,0 | 19,8 | 179 |
| Масло сливочное несоленое | 0,6 | 82,5 | 0,9 | 748 |
| Масло подсолнечное рафинированное | 0,0 | 99,9 | 0,0 | 899 |
| Крупа манная | 11,3 | 0,7 | 71,6 | 327 |
| Овсяные хлопья "Геркулес" | 13,1 | 6,2 | 62,5 | 345 |
| Рис | 7,0 | 0,5 | 74,5 | 324 |
| Крупа гречневая (ядрица) | 12,6 | 3,3 | 65,0 | 328 |
| Крупа пшенная | 12,0 | 2,8 | 66,5 | 327 |
| Крупа перловая | 9,3 | 1,1 | 67,3 | 307 |
| Крупа кукурузная | 8,3 | 1,2 | 72,4 | 325 |
| Горох лущеный | 23,0 | 1,6 | 50,8 | 287 |
| Фасоль | 22,3 | 1,7 | 44,9 | 262 |
| Соя | 34,9 | 17,3 | 9,16 | 297 |
| Макаронные изделия | 10,7 | 1,3 | 69,6 | 322 |
| Хлеб ржаной | 5,51 | 1,0 | 32,08 | 154 |
| Хлеб пшеничный 1 сорта | 7,6 | 0,9 | 49,7 | 230 |
| Хлеб зерновой | 8,6 | 1,0 | 49,8 | 234 |
| Сухари из пшеничной муки | 11,2 | 1,4 | 72,4 | 336 |
| Картофель | 2,0 | 0,4 | 16,3 | 75 |
| Баклажаны | 1,2 | 0,1 | 5,1 | 25 |
| Кабачки | 0,6 | 0,3 | 4,9 | 24 |
| Капуста белокочанная | 1,8 | 0,1 | 4,7 | 25 |
| Капуста цветная | 2,5 | 0,3 | 4,5 | 28 |
| Лук репчатый | 1,4 | 0,0 | 9,1 | 41 |
| Лук зеленый | 1,3 | 0,0 | 3,5 | 18 |
| Морковь | 1,3 | 0,1 | 7,1 | 33 |
| Огурцы свежие | 0,8 | 0,0 | 2,6 | 13 |
| Помидоры свежие | 1,1 | 0,0 | 3,8 | 19 |
| Перец болгарский сладкий | 1,3 | 0,0 | 5,3 | 25 |
| Редис | 1,2 | 0,0 | 3,8 | 19 |

| Продукт | | | |
|---|---|---|---|
| Арбуз | 0,7 | 0,0 | 8,8 | 37 |
| Дыня | 0,6 | 0,0 | 9,1 | 38 |
| Апельсины | 0,9 | 0,0 | 8,1 | 35 |
| Бананы | 1,5 | 0,0 | 21,0 | 89 |
| Виноград | 0,6 | 0,0 | 16,0 | 66 |
| Груша | 0,4 | 0,0 | 9,5 | 39 |
| Яблоко | 0,4 | 0,0 | 9,8 | 40 |
| Изюм | 1,8 | 0,0 | 66,0 | 269 |
| Курага | 5,2 | 0,0 | 55,0 | 236 |
| Сок абрикосовый | 0,5 | 0,0 | 13,7 | 56 |
| Сок апельсиновый | 0,7 | 0,0 | 12,8 | 53 |
| Сок айвовый | 0,5 | 0,0 | 10,4 | 43 |
| Сок виноградный | 0,4 | 0,0 | 18,2 | 74 |
| Сок вишневый | 0,7 | 0,0 | 12,2 | 51 |
| Сок грейпфрутовый | 0,3 | 0,0 | 8,0 | 33 |
| Сок сливовый | 0,3 | 0,0 | 16,1 | 65 |
| Сок яблочный | 0,5 | 0,0 | 10,6 | 44 |
| Сок томатный | 1,0 | 0,0 | 3,3 | 16 |
| Сахар-песок | 0,0 | 0,0 | 99,8 | 399 |
| Мед натуральный | 0,8 | 0,0 | 80,3 | 324 |
| Вафли | 3,2 | 2,8 | 80,1 | 355 |
| Зефир | 0,8 | 0,0 | 78,3 | 316 |
| Печенье сахарное | 7,5 | 11,8 | 74,4 | 426 |
| Конфеты помадные | 3,6 | 17,6 | 67,2 | 438 |
| Шоколад | 5,4 | 35,3 | 52,6 | 544 |

При необходимости усиленно вводить в рацион (или ограничивать) определенные ингридиенты, необходимо пользоваться следующими тремя таблицами, где продукты ранжированы по доле энергоемкости соответствующего ингридиента в общей энергоемкости продукта.

Таблица 2. Доля белков в энергоемкости продуктов

| Продукт | Доля белков | Продукт | Доля белков | Продукт | Доля белков |
|---|---|---|---|---|---|
| Кальмар | 0,95 | Помидоры свежие | 0,18 | Рис | 0,06 |
| Треска | 0,90 | Морская капуста мороженая | 0,16 | Арбуз | 0,06 |
| Минтай | 0,85 | Колбаса докторская | 0,16 | Мороженое сливочное | 0,06 |
| Судак | 0,85 | Кефир жирный | 0,16 | Печенье сахарное | 0,05 |
| Творог нежирный | 0,81 | Перец болгарский сладкий | 0,16 | Бананы | 0,05 |
| Ледяная рыба | 0,66 | Сардельки говяжьи | 0,15 | Сметана 15% | 0,05 |
| Окунь морской | 0,65 | Молоко пастеризированное | 0,15 | Дыня | 0,05 |
| Камбала | 0,64 | Баклажаны | 0,14 | Сметана 20% жирности | 0,04 |
| Печень говяжья | 0,62 | Сосиски говяжьи (молочные) | 0,14 | Сок вишневый | 0,04 |
| Икра зернистая | 0,43 | Колбаса любительская вареная | 0,13 | Сок апельсиновый | 0,04 |

| Продукт | Доля | Продукт | Доля | Продукт | Доля |
|---|---|---|---|---|---|
| Творог полужирный | 0,37 | Свинина мясная | 0,13 | Сок айвовый | 0,03 |
| Кефир нежирный | 0,37 | Морковь | 0,12 | Сок яблочный | 0,03 |
| Соя | 0,35 | Крупа гречневая (ядрица) | 0,12 | Груша | 0,03 |
| Кролик 1 категории | 0,32 | Овсяные хлопья "Геркулес" | 0,11 | Шоколад | 0,03 |
| Молоко обезжиренное | 0,32 | Хлеб зерновой | 0,11 | Яблоко | 0,03 |
| Говядина 1 категории | 0,31 | Крупа пшенная | 0,11 | Виноград | 0,03 |
| Капуста цветная | 0,27 | Хлеб ржаной | 0,11 | Сок грейпфрутовый | 0,03 |
| Яйцо куриное | 0,26 | Крупа манная | 0,10 | Вафли | 0,03 |
| Фасоль | 0,26 | Лук репчатый | 0,10 | Сок абрикосовый | 0,03 |
| Куры 1 категории | 0,25 | Сухари из пшеничной муки | 0,10 | Конфеты помадные | 0,02 |
| Сыр голландский | 0,25 | Макаронные изделия | 0,10 | Изюм | 0,02 |
| Баранина 1 категории | 0,24 | Хлеб пшеничный 1 сорта | 0,10 | Сметана 35% жирности | 0,02 |
| Горох лущеный | 0,24 | Крупа перловая | 0,09 | Сок виноградный | 0,02 |
| Лук зеленый | 0,22 | Сметана 10% | 0,08 | Сок сливовый | 0,01 |
| Капуста белокочанная | 0,22 | Картофель | 0,08 | Зефир | 0,01 |
| Творог жирный | 0,19 | Апельсины | 0,08 | Мед натуральный | 0,01 |
| Редис | 0,19 | Крупа кукурузная | 0,08 | Масло сливочное несоленое | 0,00 |
| Огурцы свежие | 0,19 | Кабачки | 0,07 | Масло подсолнечное рафинированное | 0,00 |
| Сок томатный | 0,19 | Курага | 0,07 | Сахар-песок | 0,00 |

Таблица 3. Доля жиров в энергоемкости продуктов

| Продукт | Доля жиров | Продукт | Доля жиров | Продукт | Доля жиров |
|---|---|---|---|---|---|
| Масло подсолнечное рафинированное | 1,00 | Ледяная рыба | 0,34 | Молоко обезжиренное | 0,02 |
| Масло сливочное несоленое | 0,99 | Печенье сахарное | 0,25 | Рис | 0,01 |
| Сметана 35% жирности | 0,95 | Овсяные хлопья "Геркулес" | 0,16 | Лук репчатый | 0,00 |
| Сметана 20% жирности | 0,89 | Судак | 0,15 | Лук зеленый | 0,00 |
| Свинина мясная | 0,87 | Минтай | 0,15 | Огурцы свежие | 0,00 |
| Колбаса любительская вареная | 0,87 | Кабачки | 0,11 | Помидоры свежие | 0,00 |
| Сосиски говяжьи (молочные) | 0,86 | Морская капуста мороженая | 0,11 | Перец болгарский сладкий | 0,00 |
| Сметана 15% | 0,86 | Треска | 0,10 | Редис | 0,00 |
| Колбаса докторская | 0,84 | Капуста цветная | 0,10 | Арбуз | 0,00 |
| Сметана 10% | 0,81 | Крупа гречневая (ядрица) | 0,09 | Дыня | 0,00 |
| Сардельки говяжьи | 0,81 | Творог нежирный | 0,08 | Апельсины | 0,00 |
| Баранина 1 категории | 0,76 | Крупа пшенная | 0,08 | Бананы | 0,00 |

| Сыр голландский | 0,75 | Вафли | 0,07 | Виноград | 0,00 |
| Творог жирный | 0,75 | Хлеб ржаной | 0,06 | Груша | 0,00 |
| Куры 1 категории | 0,75 | Фасоль | 0,06 | Яблоко | 0,00 |
| Яйцо куриное | 0,72 | Горох лущеный | 0,05 | Изюм | 0,00 |
| Говядина 1 категории | 0,69 | Картофель | 0,05 | Курага | 0,00 |
| Кролик 1 категории | 0,68 | Кальмар | 0,05 | Сок абрикосовый | 0,00 |
| Творог полужирный | 0,59 | Хлеб зерновой | 0,04 | Сок апельсиновый | 0,00 |
| Шоколад | 0,58 | Сухари из пшеничной муки | 0,04 | Сок айвовый | 0,00 |
| Икра зернистая | 0,57 | Макаронные изделия | 0,04 | Сок виноградный | 0,00 |
| Кефир жирный | 0,54 | Баклажаны | 0,04 | Сок вишневый | 0,00 |
| Соя | 0,52 | Капуста белокочанная | 0,04 | Сок грейпфрутовый | 0,00 |
| Молоко пастеризированное | 0,51 | Хлеб пшеничный 1 сорта | 0,04 | Сок сливовый | 0,00 |
| Мороженое сливочное | 0,50 | Крупа кукурузная | 0,03 | Сок яблочный | 0,00 |
| Печень говяжья | 0,38 | Крупа перловая | 0,03 | Сок томатный | 0,00 |
| Камбала | 0,36 | Морковь | 0,03 | Сахар-песок | 0,00 |
| Конфеты помадные | 0,36 | Крупа манная | 0,02 | Мед натуральный | 0,00 |
| Окунь морской | 0,35 | Кефир нежирный | 0,02 | Зефир | 0,00 |

Таблица 4. Доля углеводов в энергоемкости продуктов

| Продукт | Доля углеводов | Продукт | Доля углеводов | Продукт | Доля углеводов |
|---|---|---|---|---|---|
| Сахар-песок | 1,00 | Сухари из пшеничной муки | 0,86 | Творог нежирный | 0,11 |
| Мед натуральный | 0,99 | Морковь | 0,86 | Сметана 10% | 0,10 |
| Зефир | 0,99 | Хлеб зерновой | 0,85 | Сметана 15% | 0,09 |
| Сок сливовый | 0,99 | Перец болгарский сладкий | 0,84 | Сметана 20% жирности | 0,06 |
| Сок виноградный | 0,98 | Хлеб ржаной | 0,83 | Творог жирный | 0,05 |
| Изюм | 0,98 | Помидоры свежие | 0,82 | Сардельки говяжьи | 0,04 |
| Сок абрикосовый | 0,97 | Баклажаны | 0,82 | Творог полужирный | 0,04 |
| Виноград | 0,97 | Сок томатный | 0,81 | Сметана 35% жирности | 0,03 |
| Сок грейпфрутовый | 0,97 | Кабачки | 0,81 | Яйцо куриное | 0,02 |
| Яблоко | 0,97 | Крупа пшенная | 0,81 | Масло сливочное несоленое | 0,00 |
| Груша | 0,97 | Огурцы свежие | 0,81 | Говядина 1 категории | 0,00 |
| Сок яблочный | 0,97 | Редис | 0,81 | Печень говяжья | 0,00 |
| Сок айвовый | 0,97 | Крупа гречневая (ядрица) | 0,79 | Баранина 1 категории | 0,00 |
| Сок апельсиновый | 0,96 | Лук зеленый | 0,78 | Свинина мясная | 0,00 |
| Сок вишневый | 0,96 | Капуста белокочанная | 0,75 | Кролик 1 категории | 0,00 |
| Дыня | 0,95 | Морская капуста мороженая | 0,73 | Куры 1 категории | 0,00 |
| Бананы | 0,95 | Овсяные хлопья | 0,72 | Сосиски говяжьи | 0,00 |

| | | "Геркулес" | | (молочные) | |
|---|---|---|---|---|---|
| Арбуз | 0,94 | Горох лущеный | 0,71 | Колбаса любительская вареная | 0,00 |
| Курага | 0,93 | Печенье сахарное | 0,70 | Колбаса докторская | 0,00 |
| Апельсины | 0,92 | Фасоль | 0,69 | Камбала | 0,00 |
| Рис | 0,92 | Молоко обезжиренное | 0,67 | Ледяная рыба | 0,00 |
| Вафли | 0,90 | Капуста цветная | 0,64 | Минтай | 0,00 |
| Лук репчатый | 0,90 | Кефир нежирный | 0,62 | Окунь морской | 0,00 |
| Крупа кукурузная | 0,89 | Конфеты помадные | 0,61 | Судак | 0,00 |
| Крупа манная | 0,88 | Мороженое сливочное | 0,44 | Треска | 0,00 |
| Крупа перловая | 0,88 | Шоколад | 0,39 | Икра зернистая | 0,00 |
| Картофель | 0,87 | Молоко пастеризированное | 0,34 | Кальмар | 0,00 |
| Хлеб пшеничный 1 сорта | 0,87 | Кефир жирный | 0,31 | Сыр голландский | 0,00 |
| Макаронные изделия | 0,86 | Соя | 0,12 | Масло подсолнечное рафинированное | 0,00 |

Необходимо также изменить процентное соотношение между белками, жирами и углеводами в суточной энергоемкости рациона. Так, общепринятое соотношение для здоровых людей молодого возраста, живущих в умеренном климате и не занятых физическим трудом (белки – 13%, жиры – 33%, углеводы – 54%), исходя из СДДП белков, следует изменить на (белки – 10%, жиры – 34%, углеводы – 56%). При этом подчеркиваем, что изменяются не абсолютные количества (в граммах), белков, жиров и углеводов, а их стоимостное выражение в ЦЕП исходя из СДДП по сравнению со стоимостным выражением в килокалориях. Эти границы (10% для белков, 34% для жиров и 56% для углеводов) отражены в таблицах 2–4 горизонтальной линией. Лежащие в таблицах выше горизонтальной линии продукты могут рассматриваться как поставщики соответствующего ингридиента при необходимости увеличения его доли в рационе, и их необходимо исключить из рациона при соответствующей установке в питании на уменьшение доли соответствующего ингридиента. Продукты, лежащие в таблицах ниже горизонтальной линии, являются обедненными в указанном отношении, и рацион насыщается ими при необходимости уменьшить долю соответствующего ингридиента в питании.

**Литература**